\documentclass{pasj00}

\SetRunningHead{S.~Matsushita and H.~Matsuo}
	{Submm Opacity at Pampa la Bola III}

\title{FTS Measurements of Submillimeter-Wave Atmospheric Opacity
	at Pampa la Bola\\
	III. Water Vapor, Liquid Water, and 183~GHz Water Vapor Line
	Opacities}

\author{Satoki \textsc{Matsushita}}
\affil{Harvard-Smithsonian Center for Astrophysics,
	60 Garden Street, MS-78, Cambridge, MA 02138, USA}
\email{smatsushita@cfa.harvard.edu}
\and
\author{Hiroshi \textsc{Matsuo}}
\affil{Advanced Technology Center, National Astronomical Observatory
	of Japan, Mitaka, Tokyo 181-8588}
\email{h.matsuo@nao.ac.jp}

\Received{2002 August 12}
\Accepted{2002 November 16}
\Published{}

\begin{document}

\maketitle

\begin{abstract}
Further analysis has been made on the millimeter and
submillimeter-wave (100--1600~GHz or 3~mm -- 188~$\mu$m) atmospheric
opacity data taken with the Fourier Transform Spectrometer (FTS) at
Pampa la Bola, 4800~m above sea level in northern Chile, which is
the site of the Atacama Large Millimeter/submillimeter Array (ALMA).
Time-sequence plots of millimeter and submillimeter-wave opacities
show similar variations to each other, except for during the periods
with liquid water (fog or clouds) in the atmosphere.
Using millimeter and submillimeter-wave opacity correlations under
two conditions, which are affected and not affected by liquid water,
we succeeded to separate the measured opacity into water vapor and
liquid water opacity components.
The water vapor opacity shows good correlation with the 183~GHz water
vapor line opacity, which is also covered in the measured spectra.
On the other hand, the liquid water opacity and the 183~GHz line
opacity show no correlation.
Since only the water vapor component is expected to affect the phase
of interferometers significantly, and the submillimeter-wave opacity
is less affected by the liquid water component,
it may be possible to use the submillimeter-wave opacity for
a phase-correction of submillimeter interferometers.
\end{abstract}

\section{Introduction}
\label{sect-intro}

In astronomy, observations with millimeter and submillimeter
wavelengths are highly suitable for studying molecular gas and dust
emission from various astronomical objects, including young stellar
objects, star-forming/starburst regions, and primeval galaxies.
However, observations at these frequencies/wavelengths, especially at
submillimeter-wavelength, are difficult, since it is easily and
heavily absorbed by water constituents in the atmosphere.
Therefore, understanding the absorption/emission spectra caused by
the water constituents is important, but is not yet well
characterized.

At the high-altitude site of the Atacama desert in Northern Chile
(Pampa la Bola -- Chajnantor region, about 5000~m above the sea
level), the US, Europe, and Japan are planning to build the Atacama
Large Millimeter/submillimeter Array (ALMA), which will be
the world's largest millimeter and submillimeter interferometer.
Since the site is very dry, it is a suitable location for studying
the behavior of water constituents in the millimeter/submillimeter
region.
At Pampa la Bola -- Chajnantor area, 220~GHz/225~GHz tipping
radiometers \citep{koh95,rad98,rad00,rad01,rad02,sak02}, a 350~$\mu$m
tipping photometer \citep{rad98,rad02}, 183~GHz water line
radiometers (\cite{del99}, \yearcite{del00}, \yearcite{del01}), and
11~GHz radio seeing monitors
\citep{rad96,hol97,rad98,but01,rad02,sak02} are now continuously
operating to evaluate the absorption and phase fluctuation caused by
water constituents.
Also, a 492~GHz tipping radiometer performed measurements for
intermittent periods of time \citep{hir98}.
We then carried out wide frequency coverage opacity measurements
using a Fourier Transform Spectrometer (FTS) in the winter of 1997
and 1998.

Opacity measurements with the FTS were performed during 1997
September 5 and 12.
Opacity correlations between 220~GHz and all the submillimeter-wave
windows were obtained, although the weather conditions were limited
(\cite{mat98a},b).
During 1998 June 16 and 18, opacities under good weather conditions
were measured, and high transmission spectra as well as the detection
of supra-terahertz windows located around 1035~GHz, 1350~GHz, and
1500~GHz have been obtained (\cite{mat99}, \yearcite{mat00};
\cite{mat02b}).
These supra-terahertz windows have also been detected by other groups
(\cite{pai00,par01b}, \yearcite{par02}) and successfully modeled with
radiative transfer calculations (\cite{par01a},b, \yearcite{par02};
\cite{mat99}, \yearcite{mat00}).
Combining these data sets with long-term site testing radiometer data
\citep{rad98}, the fraction of time with a low-opacity condition in
submillimeter-wave (675~GHz and 875~GHz) atmospheric windows at
the ALMA site was estimated.
The submillimeter zenith opacity is less than 1.0 for about 50\% of
the winter season and about 37\% of a year, which is about 50\%
better than the Mauna Kea site \citep{mat99}.

During the measurement period in 1998, quite different opacity
correlations compared with the normal ones mentioned above were
obtained.
Because of their behavior, weather condition, and theoretical
perspectives, it is suggested that liquid water in the atmosphere
caused these correlation changes \citep{mat99,mat02b}.
We analyzed these `abnormal' opacity correlation data in detail, and
considered the effect of liquid water on millimeter- and
submillimeter-waves.
Based on these results, we propose new phase-correction methods for
submillimeter interferometry.

\section{Measurements and Data Analysis}
\label{sect-meas}

Atmospheric opacity measurements were performed at Pampa la Bola
(northern Chile, the Atacama desert, 4800~m altitude) using
a Martin-Puplett type FTS \citep{mar70} and an InSb bolometer.
The frequency coverage was 100--1600~GHz
(or 3~mm to 188~$\mu$m in the wavelength range).
Two different observing runs were carried out, the first one in 1997
September 5--12, and the second one in 1998 June 16--18
(both in winter season in Chile).

The beam size of the instrument was about \timeform{10D}, and
the apodized frequency resolution was $\sim$ 0.3~cm$^{-1}$
($\sim$ 10~GHz).
Atmospheric emission spectra were obtained toward different air
masses
[$\sec(z)$, $z$ is zenith angle] at 1.0, 1.5, 2.0, 2.5, and 3.0 by
rotating a tipping mirror outside of the spectrometer.
A tipping measurement with the FTS including two temperature
calibrations was made every 12~min in the first observing run and
every 14~min in the second run.
The absolute-brightness temperature of the atmosphere was calibrated
using blackbodies (Eccosorb AN74) at liquid-nitrogen temperature
($\sim$ 73~K at 4800~m altitude) and at ambient temperature.
The ambient temperature was monitored with a weather station
\citep{sak00}.
In data analysis, tipping measurements were used for frequencies
lower than 450~GHz, and temperature measurements were used for
frequencies higher than 450~GHz.
From 10~PM on June 17 to 4~AM on June 18 in 1998, however,
temperature measurements were used for all frequencies.
This is because the existence of clouds in the atmosphere is
suggested (see subsection~\ref{sect-abn}), and therefore requirements
for tipping measurement (need a constant sky condition toward
different air masses and a low opacity condition) would not be
satisfied.
Further details on the instruments, measurements, calibrations, and
data analysis are described elsewhere
(\cite{mat98a},b; \cite{mat99}).

Since the frequency coverage of our FTS data ranges from 100~GHz,
we can measure the 183~GHz water vapor line opacity by subtracting
the continuum component from both sides of the line.
We assumed 161--204~GHz as the 183~GHz water vapor line frequency
range, and 136--161~GHz and 204--230~GHz as continuum frequency
ranges.
The continuum opacity within the line frequency range was calculated
with the interpolation of the continuum opacities located at both
sides of the line.
After continuum subtraction, the line opacity was divided by
the number of the data points within the line frequency range
(12 points), since our FTS frequency resolution is not good enough to
resolve the 183~GHz line.
The 183~GHz opacity used in this paper is, therefore, the average
opacity in this frequency range.

\begin{figure*}
\begin{center}
\FigureFile(170mm,170mm){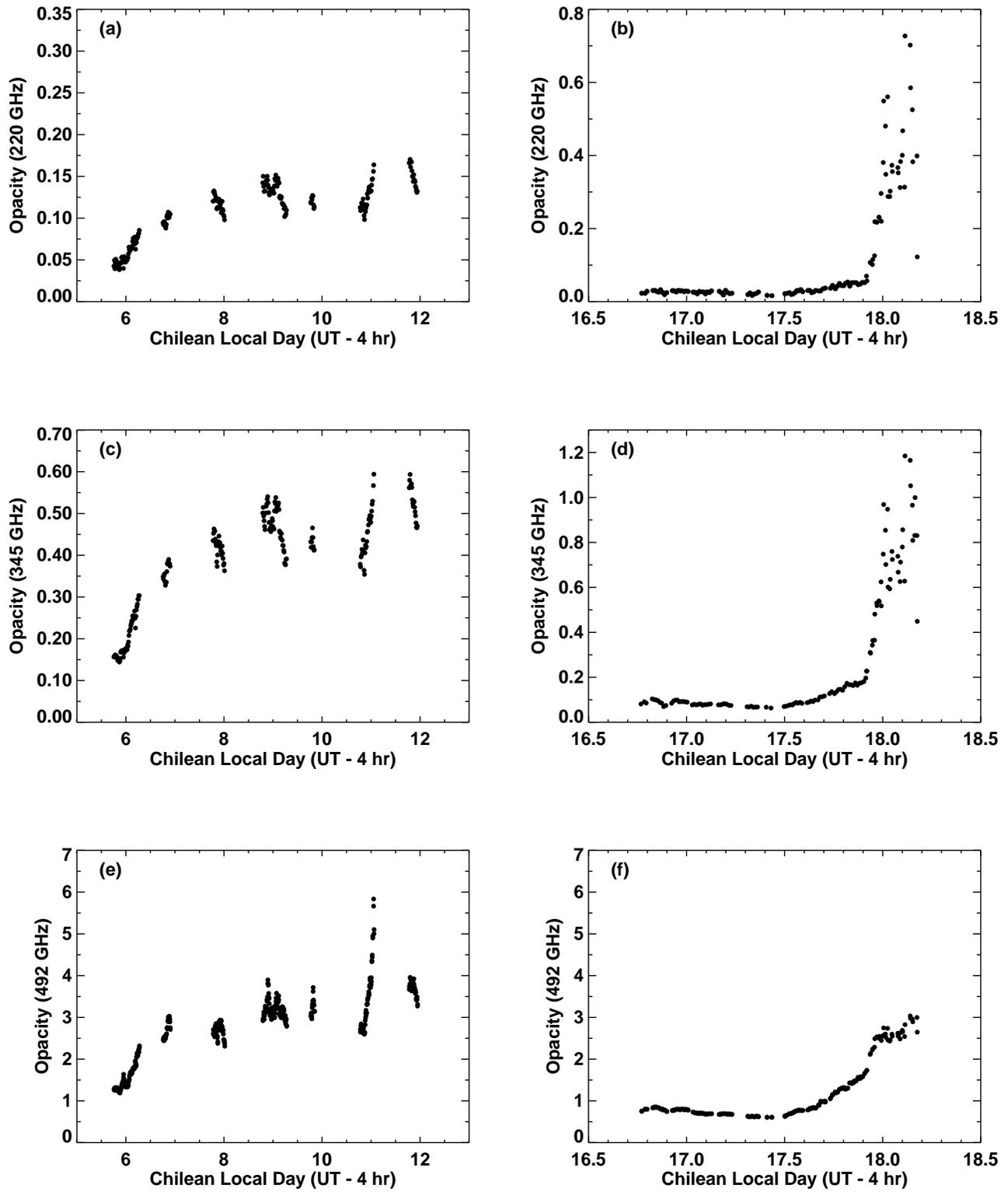}
\end{center}
\caption{Time-sequence plots for 220, 345, 492, 675, 875, and 937~GHz
	opacities.
	The left-hand and right-hand sides of the figure are the data
	taken in 1997 September and 1998 June, respectively.
	The Chilean local time corresponds to UT $-$ 4 hr.
	}
\label{time-seq1}
\end{figure*}

\addtocounter{figure}{-1}
\begin{figure*}
\begin{center}
\FigureFile(170mm,170mm){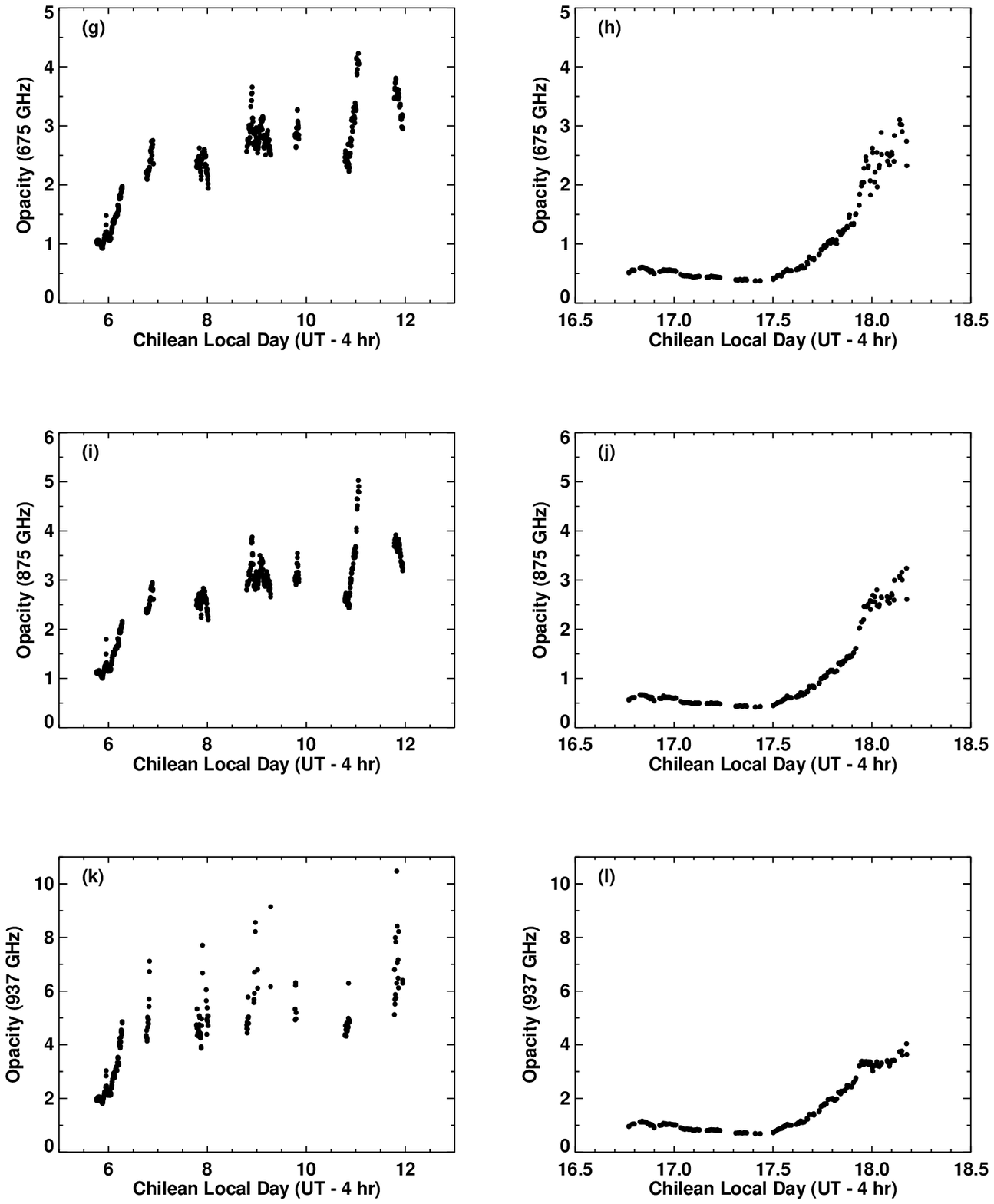}
\end{center}
\caption{(Continued)}
\end{figure*}

\begin{figure}
\begin{center}
\FigureFile(80mm,80mm){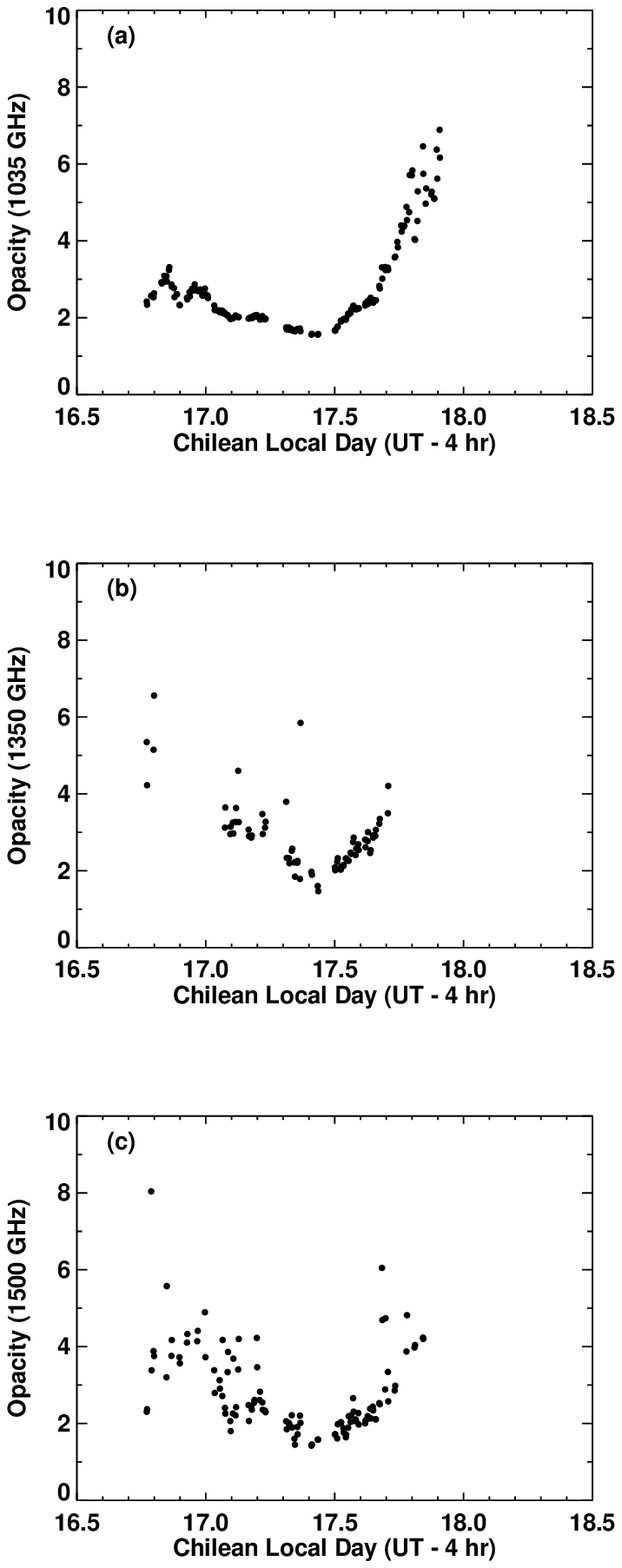}
\end{center}
\caption{Time-sequence plots for (a) 1035~GHz, (b) 1350~GHz, and
	(c) 1500~GHz opacities.
	The Chilean local time corresponds to UT $-$ 4 hr.
	Since the weather conditions during the 1997 measurement were not
	good enough to detect these supra-terahertz windows, we only show
	1998 data plots.
	}
\label{time-seq2}
\end{figure}

\begin{figure}
\begin{center}
\FigureFile(80mm,80mm){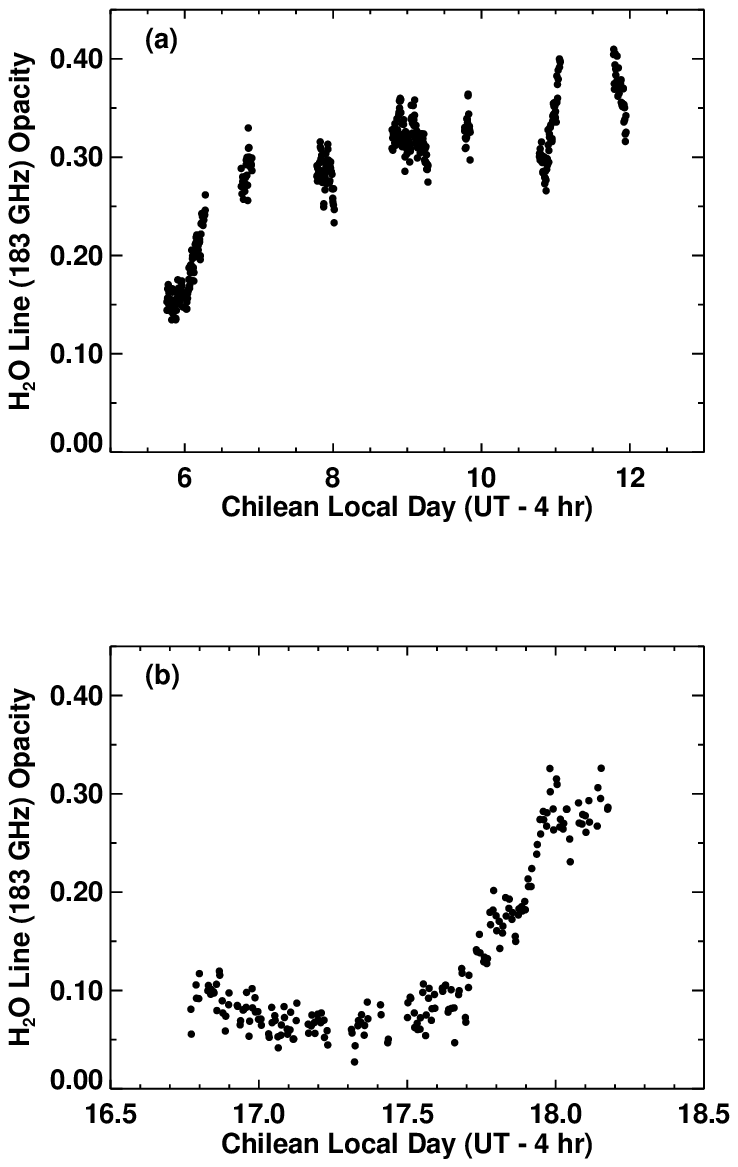}
\end{center}
\caption{Time-sequence plots for 183~GHz water vapor line opacity
	averaged over 161 to 204~GHz, taken in (a) 1997 September and
	(b) 1998 June.
	The Chilean local time corresponds to UT $-$ 4 hr.
	}
\label{time-seq-h2o}
\end{figure}

\section{Time-Variations of Measured Opacities}
\label{sect-time}

Time-variation plots for 220, 345, 492, 675, 875, and 937~GHz
opacities are shown in figure~\ref{time-seq1}.
The selected frequencies are either located at the center of
atmospheric windows and/or are frequencies of special astronomical
interests (see \cite{mat99,mat02b}).
The left-hand and right-hand sides of the figure are the data taken
in 1997 and 1998, respectively.
In the 1997 measurement, because the sensitivity of the bolometer in
daytime data had large variations, we only used nighttime data.

The time-variation plots of the 1997 data (the left-hand side of
figure~\ref{time-seq1}) clearly show that the opacity condition was
relatively good at the beginning of the measurement (1997 September
5), but gradually worsened toward the end (1997 September 12).
The overall time variations of the plots look similar to each other,
except for the plot of 937~GHz which saturates under worse weather
conditions.

On the other hand, the time-variation plots of the 1998 data
(the right-hand side of figure~\ref{time-seq1}) indicate that
the opacity condition was extremely good during the first half of
the measurement (from the evening of June 16 to around noon of 17),
but gradually worsened from the afternoon of June 17, and became
totally bad around midnight.
The lowest opacity (the best atmospheric transmission; see figure 1
in \cite{mat99}) was recorded around noon on June 17 (Chilean local
day of around 17.5 in the plots).
The first half of the time-variation plots looks similar for all
frequencies (from 220~GHz to 937~GHz), but the second half,
especially around the beginning of June 18, of the plots is different
between the lower frequencies (220~GHz and 345~GHz) and higher
frequencies.
The implications of this is discussed in a later section.

Figure~\ref{time-seq2} shows time-variation plots of
the supra-terahertz windows.
Since the weather condition of the 1997 measurement was not good
enough to detect these supra-terahertz windows, we only show the 1998
data.
Because the sensitivity of our FTS around the 1350~GHz and 1500~GHz
windows is low and has some systematic errors (see \cite{mat99} for
detail), the plots have a larger scatter than the other plots.
Still, the plots show a similar trend to the others.

We also plot the time variation of the 183~GHz water vapor line
opacity (figure~\ref{time-seq-h2o}).
The 1997 data plot (figure~\ref{time-seq-h2o}a) shows similarities to
the other opacity plots (the left-hand side of
figure~\ref{time-seq1}).
The 1998 data plot (figure~\ref{time-seq-h2o}b) shows similar
variation to the higher frequency ($\ge$ 492~GHz) plots (the right-hand
side of figure~\ref{time-seq1} and figure~\ref{time-seq2}), but
different from the lower frequency (220~GHz and 345~GHz) plots,
especially from the late evening of June 17 to the early morning of
June 18.

\section{Analysis of Water Vapor and Liquid Water Opacity Components}
\label{sect-ana}

\subsection{`Abnormal' Opacity Correlations between Millimeter- and
	Submillimeter-Wave}
\label{sect-abn}

Usually, correlation plots between 220~GHz opacity and submillimeter
window opacities are strongly correlated with coefficients of 22--24
(hereafter we refer to these correlations as `normal' correlations;
\cite{mat98a,mat99}, \yearcite{mat00}; see also \cite{mas94}).
However, from 17.9 to 18.2 Chilean local day (from 10~PM on June 17
to 4~AM on June 18) in 1998, we obtained quite different correlation
plots from the `normal' correlations (see figure~4 in \cite{mat99}).
These `abnormal' plots show lower correlation coefficients and larger
zero-point offsets than those of the `normal' correlations, which
means that the fractional variation of the millimeter opacity is much
larger than the submillimeter opacity.

The NRAO surveillance camera at the Chajnantor site
(http://www.tuc.nrao.edu/alma/site/site.html)
shows a clear sky image around 5~PM on June 17, but shows a cloudy
image around 9~AM on June 18.
In addition, the ambient temperature data taken by the weather
station \citep{sak00} recorded temperature below freezing (around
$-5\arcdeg$C to $-8\arcdeg$C).
These weather data suggest that there were water droplets (liquid
water), or ice particles, in the atmosphere when we obtained
the `abnormal' correlation data.

Water vapor absorbs stronger in the submillimeter-wave region than in
the millimeter-wave region \citep{lie89,lie93,par01a}.
Liquid water (cloud or fog), on the other hand, absorbs more
effectively at the millimeter-wave region for a given particle-size
distribution (\cite{ray72,lie89,lie91}; see also \cite{mat02a}),
which means that a variation in the quantity of the liquid water
content in the atmosphere causes a large change in
the millimeter-wave opacity.
The existence of ice cannot explain the `abnormal' correlations.
Ice has about two orders of magnitude lower values for the imaginary
part of the index of refraction ($\approx$ absorption
coefficient\footnotemark) at millimeter-wave region than that of
liquid water \citep{war84,lie93,huf91,ray72}, and is similar to that
of water vapor.
\footnotetext{The imaginary part of the complex index of refraction,
  $n_{\rm I}$, and the absorption coefficient, $\alpha$, is related
  as $\alpha=4\pi\nu n_{\rm I}/c$ \citep{tho01}.}

There is a possibility that other molecules or aerosols cause
the `abnormal' correlations.
Neither the oxygen nor ozone lines can be the cause, since these show
up with narrow lines in the spectra (e.g., \cite{par01a,mat99}).
The characteristics of the wet and dry pseudocontinuum components,
which are thought to be the far wings of water vapor lines in
the infrared range and collision-induced absorption of the dry
atmosphere (N$_{2}$ and O$_{2}$), respectively (\cite{par01a},b,
\yearcite{par02}), are still empirical and not well understood.
Furthermore, the wet pseudocontinuum should be correlated with water
vapor, and the dry atmosphere should have a constant density and is
not expected to fluctuate.
In addition, since the wet and dry pseudocontinuum absorptions at
millimeter-wave are too weak to explain the `abnormal' correlations,
we have to introduce a new kind of pseudocontinuum absorption
component, which has rather stronger absorption in
the millimeter-wave range than the other pseudocontinuum components.
However, it is beyond the scope of our paper to introduce the new
pseudocontinuum component.

We therefore suggest that the `abnormal' correlations are caused by
the effect of the liquid water absorption, which is the most
plausible known component.
The reason why liquid water still existed in the atmosphere even when
the ambient temperature was below freezing may be that
the atmospheric temperature at the clouds was warmer than that on
the ground, or that the clouds consisted of both liquid water and
ice, but the absorption caused by liquid water dominated the total
atmospheric absorption.

\subsection{Separation of Total Opacity into Two Opacity Components}
\label{sect-sep}

To confirm this suggestion, we separate the total opacity into water
vapor opacity and liquid water opacity components using the `normal'
and `abnormal' correlations between millimeter and submillimeter
opacities.
For the separation, we assume that
(a) the `normal' correlations result only from the water vapor
	absorption,
(b) `abnormal' correlations are the combined effect of the water
	vapor and liquid water absorptions,
(c) the correlation coefficients of the liquid water absorption
	correspond to those of the `abnormal' correlations, and
(d) the zero-point offsets are zero in all the correlations.
Under these assumptions, we can write:
\begin{equation}
\tau_{\nu,{\rm WV}} = a_{\rm WV} \cdot \tau_{\rm 220, WV},
\end{equation}
\begin{equation}
\tau_{\nu,{\rm LQ}} = a_{\rm LQ} \cdot \tau_{\rm 220, LQ},
\end{equation}
\begin{equation}
\tau_{\rm 220} = \tau_{\rm 220, WV} + \tau_{\rm 220, LQ},
\end{equation}
\begin{equation}
\tau_{\nu} = \tau_{\nu,{\rm WV}} + \tau_{\nu,{\rm LQ}},
\end{equation}
where $\tau_{220}$ and $\tau_{\nu}$ indicate opacities at 220~GHz and
frequency $\nu$, and the suffix WV and LQ indicate water vapor and
liquid water components, respectively.
The known (measured) parameters are $a_{\rm WV}$, $a_{\rm LQ}$,
$\tau_{\rm 220}$, and $\tau_{\nu}$.
Using these parameters, we can separate the water vapor and liquid
water opacity components from the measured total opacity.
We summarized the $a_{\rm WV}$ and $a_{\rm LQ}$ of the selected
frequencies derived from the `normal' \citep{mat00} and `abnormal'
\citep{mat99} correlations in table~\ref{tab-awv-alq}.

\begin{table}
\begin{center}
\caption{Correlation coefficients of the `normal' ($a_{\rm WV}$) and
	`abnormal' ($a_{\rm LQ}$) correlations between 220 GHz opacity
	and submillimeter opacities.}
\label{tab-awv-alq}
\begin{tabular}{ccc}
\hline\hline
Frequency &  $a_{\rm WV}$\footnotemark[$*$] &  $a_{\rm LQ}$ \\
  (GHz)   &               &               \\ \hline
   345    & $3.65\pm0.02$ & $1.42\pm0.03$ \\
   410    & $7.53\pm0.05$ & $1.64\pm0.05$ \\
   492    &  $23.6\pm0.3$ &  $1.5\pm0.2$  \\
   675    &  $22.4\pm0.2$ &  $1.9\pm0.2$  \\
   875    &  $24.2\pm0.2$ &  $1.8\pm0.2$  \\
   937    &  $43.9\pm0.9$ &  $0.6\pm0.2$  \\ \hline
\multicolumn{3}{@{}l@{}}{\hbox to 0pt{\parbox{85mm}{\footnotesize
\par\noindent
\footnotemark[$*$] The values are taken from \citet{mat00}.
}\hss}}
\end{tabular}
\end{center}
\end{table}

\begin{figure*}
\begin{center}
\FigureFile(170mm,170mm){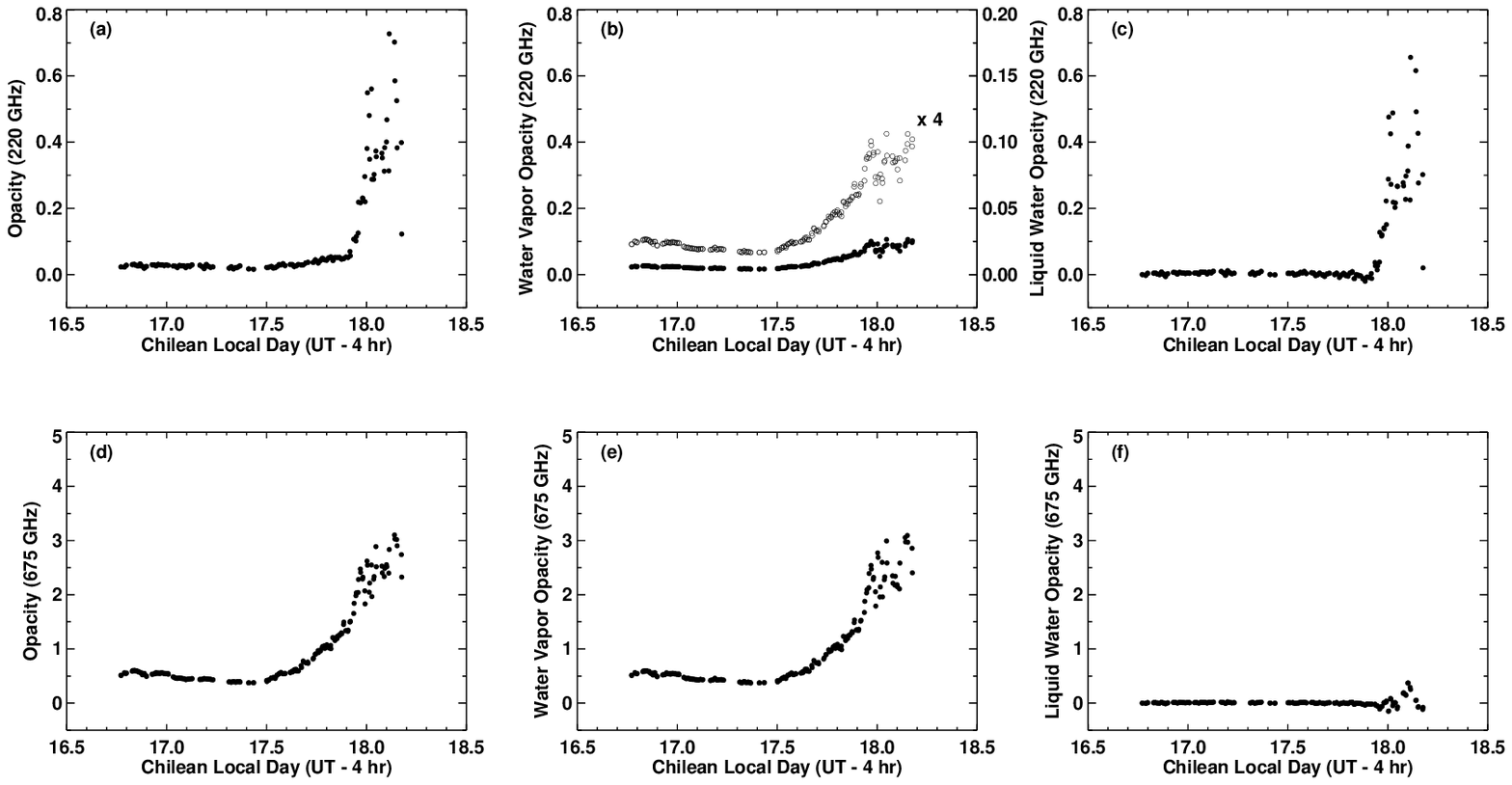}
\end{center}
\caption{Time variations of
	(a) total 220~GHz opacity,
	(b) 220~GHz separated water vapor opacity (filled circles, see
	left-hand axis for their unit), overplotted with the same plot
	but multiplied by four (open circles, see right-hand axis),
	(c) 220~GHz separated liquid water opacity,
	(d) total 675~GHz opacity,
	(e) 675~GHz separated water vapor opacity, and
	(f) 675~GHz separated liquid water opacity.
	The data were taken during the 1998 measurement, and the Chilean
	local time corresponds to UT $-$ 4 hr.}
\label{time-seq-sep}
\end{figure*}

We then performed the component separation on the 220~GHz and 675~GHz
opacities taken in 1998,
and the results of the separated millimeter and submillimeter
opacities are shown in figure~\ref{time-seq-sep}.
The top row of figure~\ref{time-seq-sep} shows the time variations at
220~GHz for (a) the total opacity, (b) the separated water vapor
opacity, and (c) the separated liquid water opacity.
The bottom row shows the same plots but for 675~GHz.
The total opacity plots for 220~GHz and 675~GHz
(figures~\ref{time-seq-sep}a,d) look different, especially around
Chilean local day from 17.9 to 18.2, but the separated water vapor
opacity plots (figures~\ref{time-seq-sep}b,e) show similar time
variation to each other.
On the other hand, the 220~GHz separated liquid water opacity plot
(figure~\ref{time-seq-sep}c) shows a quite different behavior from
that of 675~GHz (figure~\ref{time-seq-sep}f), when the `abnormal'
correlation occurred.
These results suggest that millimeter opacity is largely affected by
the liquid water component in the atmosphere, although the effect in
submillimeter opacity is small.

\subsection{Correlations with 183 GHz Water Vapor Line Opacity}
\label{sect-corr}

\begin{figure*}
\begin{center}
\FigureFile(170mm,170mm){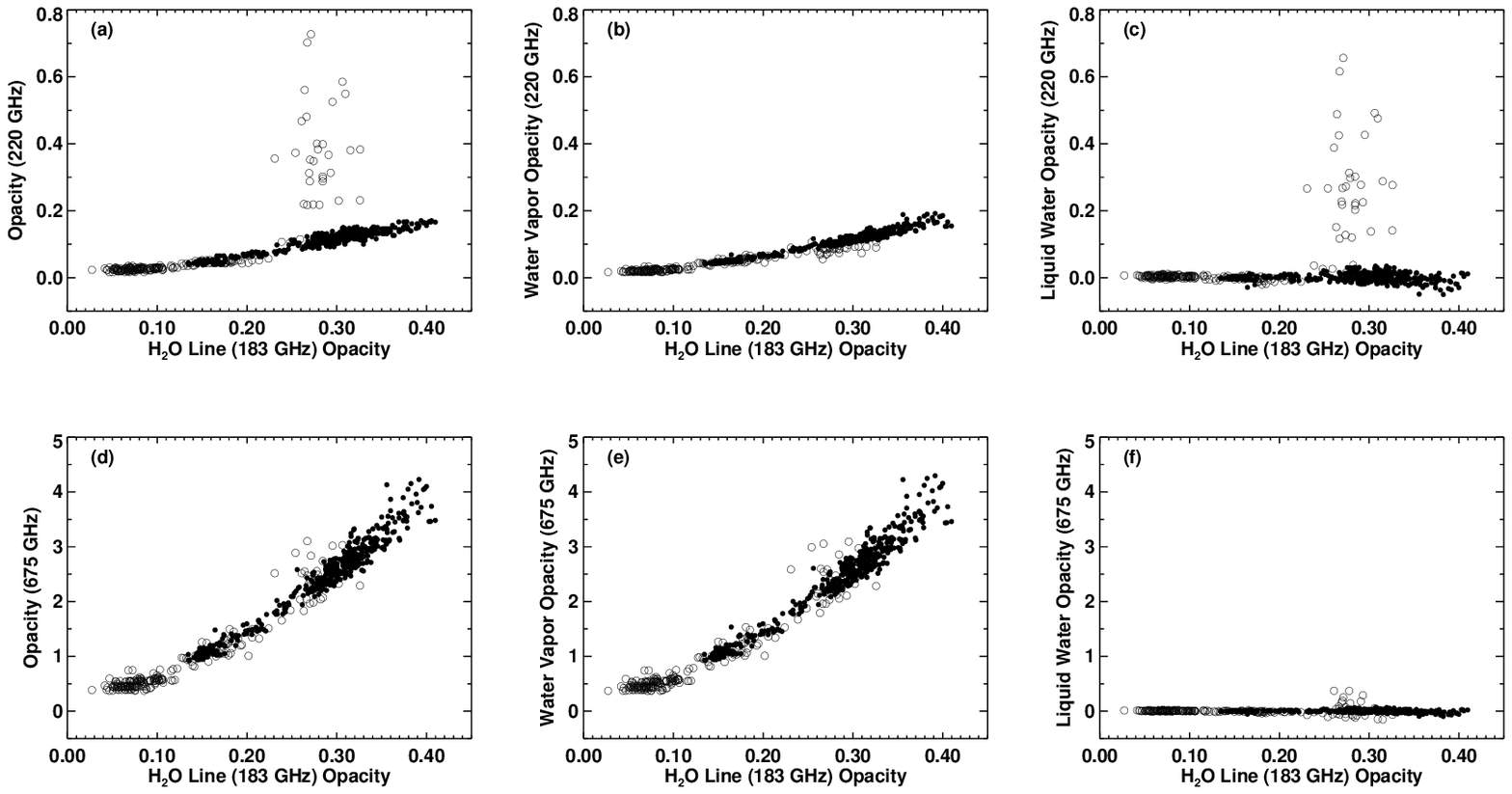}
\end{center}
\caption{Correlation diagrams between 183~GHz water-vapor line
	opacity averaged over 161 to 204~GHz and
	(a) total 220~GHz opacity,
	(b) 220~GHz separated water vapor opacity,
	(c) 220~GHz separated liquid water opacity,
	(d) total 675~GHz opacity,
	(e) 675~GHz separated water vapor opacity, and
	(f) 675~GHz separated liquid water opacity.
	The filled and open circles indicate the dataset taken in 1997
	September and 1998 June, respectively.}
\label{corr-sep}
\end{figure*}

To further confirm that the separated water vapor and liquid water
opacities were really caused by these components in the atmosphere,
it is very important to compare these with the 183~GHz water vapor
line opacity.
To this end, we made correlation diagrams between the 183~GHz line
opacity and those of total, water vapor, and liquid water at 220~GHz
(millimeter) and 675~GHz (submillimeter), as shown in
figure~\ref{corr-sep}.
The correlation diagram between the 183~GHz line opacity and 220~GHz
total opacity (figure~\ref{corr-sep}a) shows a large scatter, but
after the separation, the 220~GHz water vapor opacity shows good
correlation with the 183~GHz line opacity (figure~\ref{corr-sep}b).
On the other hand, the 220~GHz liquid water opacity is completely
independent from the 183~GHz line opacity (figure~\ref{corr-sep}c).
These results confirm the success of the opacity component
separation.

The correlation diagram of the 183~GHz line opacity and the 675~GHz
total opacity (figure~\ref{corr-sep}d) is similar to that of
the 183~GHz line opacity and the 675~GHz water vapor opacity.
The correlation diagram of the 183~GHz line opacity and the 675~GHz
liquid water opacity, on the other hand, shows little contribution of
liquid water in the 675~GHz opacity.
These results support that the submillimeter opacity is dominated by
water vapor absorption, and less affected by liquid water absorption.

The correlation diagrams between the water vapor opacity and
the 183~GHz line opacity (figures~\ref{corr-sep}b,e) show a somewhat
curved structure.
The curve at the lower end of the opacity, which shows an offset from
the zero opacity of the water vapor opacity, is caused by
the limitation of the measurements (see \cite{mat98a},b) and/or by
both the wet and dry pseudocontinuum absorptions.
The curve at the upper end can be explained by the saturation of
the 183~GHz line.
This saturation is theoretically expected \citep{lay98,car98} and
also observed with multi-channel 183~GHz line radiometers
\citep{wie00,yun99,wie01}.

\section{Application for Phase-Correction}
\label{sect-appl}

The most unwanted atmospheric effect for millimeter/submillimeter
interferometers is excess paths in electromagnetic wave propagations
caused by water vapor.
A fluctuation in the path difference between a pair of antennas
causes phase fluctuation, which decreases the interferometric data
quality noticeably.
Since the characteristics of water vapor are theoretically and
observationally well known, it is better to use its absorption for
phase-correction methods.
Liquid water, on the other hand, has a different index of refraction
from that of water vapor \citep{tho01,ray72,lie89}, and its
contribution to the opacity at lower frequency is much higher than
that of water vapor, but affects the phase fluctuations less
\citep{mat02a}.
In addition, the characteristics of liquid water are poorly
understood.
Hence, for phase-corrections using opacity (brightness temperature)
of the atmosphere, one should measure the {\it water vapor} opacity
instead of the opacity affected by {\it liquid water}.

Based on the results and discussion in the previous section, we
suggest that an accurate phase-correction for interferometers can be
made using
(a) 183~GHz water vapor line radiometers, which avoids the line
	center that can be easily saturated,
(b) submillimeter-wave radiometers whose frequency range is much less
	affected by the liquid water absorption, or
(c) two frequency (millimeter- and submillimeter-wave) radiometers
	which can separate the water vapor and liquid water components,
	as mentioned above.

The phase-correction method (a) has already been discussed and
demonstrated by many authors (\cite{wie01}, \yearcite{wie02};
\cite{del01,wie00}).

Method (b) needs either one radiometer in each antenna, or it is also
possible to use the total power output from a submillimeter receiver
for astronomical observations, and hence no additional instruments
are required.
Although the effect of liquid water in the submillimeter-wave is much
less than that in millimeter-wave, this method may be inaccurate
under a large liquid water content.
We can estimate the phase-correction errors due to the liquid water
content, assuming that it contributes to a typical 220~GHz opacity
scatter of 10\%.
The estimated errors are 19~\micron, 42~\micron, and
71~\micron\ under 220 GHz opacities ($\tau_{220}$) of 0.016 (the best
condition at the ALMA site;
\cite{mat99}), 0.036, and 0.061 (25\% and 50\% quartile of
$\tau_{220}$, respectively; \cite{rad00}), respectively.
These errors correspond to phase errors of 15\timeform{D},
34\timeform{D}, and 57\timeform{D} at 675 GHz, respectively, which
are the accuracies for this phase-correction method (see
\cite{mat02a} for detail).

Method (c) needs two radiometers in each antenna or uses the total
power outputs of both millimeter and submillimeter receivers that
operate simultaneously.
In spite of the system complexity, this method is advantageous, since
there are fewer errors due to the liquid water content.
One important question is whether $a_{\rm LQ}$ is constant or not.
The emission properties of liquid water as a function of frequency
will depend on droplet size and temperature, which are likely to
change from night to night.
More measurements under large liquid water content are needed.
However, the general trend, i.e., submillimeter measurements are less
affected by liquid water, does not change.
If the uncertainty of the liquid water model is 10\%, the error of
the 675 GHz total power phase-correction mentioned above will be
reduced by 1/10 \citep{mat02a}.

The merit of method (b) is that the field of view for
the phase-correction is exactly the same as that of the astronomical
observations.
A disadvantage of submillimeter phase-correction methods is that they
cannot be used to observe in submillimeter-wavelength under bad
weather conditions, and therefore cannot be applied to a site where
the submillimeter opacity is large.

On the other hand, in some cases, the effect of ice particles might
exist.
Since the absorption characteristics of ice particles are similar to
those of water vapor (see subsection~\ref{sect-abn}), it may be
difficult to distinguish between these two absorptions by the methods
(b) and (c).
In addition, it is not clear to which amount of ice particles usually
exist in the atmosphere.
These might be one of the uncertainties in the proposed total-power
phase-correction methods.
To clarify these problems, further measurements of atmospheric
opacity spectra due to ice particles and detailed theoretical
modelings will be needed.

To check the feasibility of the newly proposed methods (b) and (c),
test measurements with other phase-correction methods, such as
the method (a), or the fast-switching method
(\cite{hol92,hol95,car96,car97}, \yearcite{car99}; \cite{mor00})
should be performed.
The best available instrument for such tests is the Sub-Millimeter
Array (SMA; \cite{ho00}) on Mauna Kea, Hawaii, which is being
constructed by the Smithsonian Astrophysical Observatory (SAO) and
the Academia Sinica Institute of Astronomy and Astrophysics (ASIAA)
of Taiwan.
SMA has a similar observing frequency range to ALMA (180--900~GHz),
and will have a capability of dual-frequency operation.
In addition, two 183~GHz radiometers \citep{wie01} are already
installed, and improved 183~GHz radiometers \citep{hil01} are being
developed.
SMA will therefore be an ideal instrument to evaluate the proposed
total-power phase-correction methods and other various methods.

\section{Conclusion}
\label{sect-concl}

We have made further analyses on the FTS measurements of millimeter
and submillimeter atmospheric opacity taken at Pampa la Bola.
Part of the 1998 opacity data
is heavily affected by liquid water in the atmosphere, which served
to understand the absorption spectrum of liquid water.
We successfully separated the measured opacity into water vapor and
liquid water opacity components.
The separated water vapor opacity shows a good correlation with
the 183~GHz water vapor line opacity, and the liquid water opacity
shows no correlation with the line opacity.

The results also indicate that the submillimeter opacity is less
affected by liquid water than the millimeter opacity; we therefore
propose to use the submillimeter opacity for the interferometric
phase-correction by using the total-power outputs from submillimeter
receivers for astronomical observations.

\vspace*{5ex}

We would like to thank Virginia Starke for carefully reading our manuscript.
We also thank Martina C.\ Wiedner and our referee for helpful comments.
This work is partly supported by Grant-in-Aid for Scientific Research
(No.~13304015) from the Japan Society for the Promotion of Science (JSPS).
This work was supported by the Inter-Research Centers Cooperative Program
of the JSPS.

\end{document}